\documentclass[aps,twocolumn,groupedaddress]{revtex4}
\usepackage{amsmath,amssymb}
\begin{document}
\bibliographystyle{apsrev}


\title{Brownian motion approach to the ideal gas of relativistic particles}




\author{Ryszard Zygad\l{}o }
\affiliation{Marian Smoluchowski
Institute of Physics, Jagiellonian University, Reymonta 4, PL--30059 Krak\'ow,
Poland}


\date{\today}

\begin{abstract}
The relativistic generalization of a free Brownian motion theory is presented.
The global characteristics of the relaxation are {\it explicitly} found for the
velocity and momentum (stochastic) kinetics.
It is shown that the thermal corrections, to the both
relaxation times $T$ (of stationary autocorrelations)
and transient relaxation time of momentum, appear
slowing down the processes. The transient relaxation time of the velocity
does not depend {\it explicitly} on temperature,
$T(v_0)= m(v_0)/\gamma \equiv \varepsilon_0/\gamma c^2$,
and it is proportional to the initial energy of a relativistic Brownian particle.

\end{abstract}

\maketitle


During the last twenty years the Brownian motion approach has become the
most important tool for discovering and studying the spectacular phenomena,
the  {\it critical slowing down} [1], {\it noise--induced transitions} [2],
{\it stochastic resonance} [3],
{\it ratchet dynamics} [4], and {\it resonant activation} [5],
in (nonlinear) nonequilibrium systems. In the most of thermodynamical
applications the
nonlinear Langevin equation appears
as an overdamped limit of a dissipative dynamic in a
certain external potential [6]. The important conclusion of the present paper
is that, within the Markovian diffusion theory, the nonlinear stochastic
kinetic results already on the basic level of (spatially homogeneous)
ideal gases description if the relativistic particles are considered.
Particularly the relativistic free Brownian
motion theory  based on identical
general assumption as the original one
[i.e., including the interaction with
the thermal bath by a {\it systematic force}, the dissipative kinematic
friction proportional to the
velocity $-\gamma v$, and an additive {\it random force}, the thermal fluctuation,
given
by Gaussian white noise $\xi_t$]
\begin{equation}            \dot{p}_t= -\gamma v +\xi_t,
\end{equation}
where [$c=1$ is used throughout the paper]
$p= mv/\sqrt{1-v^2}$ is a relativistic momentum of a Brownian particle,
$v=p/\sqrt{p^2 +m^2}= p/
\varepsilon $, $\varepsilon = \sqrt{p^2 +m^2}$, and $ \langle\xi_t \xi_s
\rangle =2D\delta(t-s)$, is consistent both with the thermodynamical requirements of
the equilibrium Gibbs ensemble theory and with the special relativity requirements of
transformational properties of the momentum probability density distribution.

First note that the general Kramers--Fokker--Planck kinetic equation
(see, e.g., the equation (26.26) of the Ref.~[7]) of a Markovian diffusion description
simplifies to the form
\begin{equation}
\partial_t P = \partial_p L_{22}[(\partial_p H_1)P  + k_B T \partial_p P]
\end{equation}
for the spatially homogeneous
system of free noninteracting particles [of the Hamiltonian $H_1(p)$].
The kinetic coefficient $L_{22} = L_{22} (p, T)$ is in principle determined by the
properties of the interaction with the thermal bath and do not depend on $p$
for the potential interaction. For the latter case
(of the state independent diffusion) the Langevin description Eq.~(1),
and the corresponding Fokker--Planck one [8], Eq.~(2), are equivalent in view of
the identification $L_{22}=\gamma=\gamma(T)$, $D=\gamma k_B T$
(the Einstein relation, or the {\it fluctuation--dissipation theorem}), and
$\partial_p H_1 =v$. The Eq.~(2) is the most general form of the diffusion in the
momentum space leading to the Gibbs--Boltzmann stationary state $W(p) \propto
e^{-\beta H_1(p)}$ of thermal equilibrium ($\beta=1/k_B T$). The corresponding
(nonlinearly coupled) Langevin equation (in the Stratonovich interpretation [8])
is
\begin{equation}            \dot{p}_t= -\gamma g^2 v +Dgg^\prime +g\xi_t,
\end{equation}
where $L_{22}= \gamma g^2$ and
$g^\prime = \partial g/\partial p$.

The (any) corresponding probability density distributions
$W(p)$ and $W^\prime (p^\prime)$
in a resting and moving (with a constant velocity $V$) reference frame
are related by non-covariant formula
$\varepsilon W(p) = \varepsilon^\prime W^\prime (p^\prime)$,
where $\varepsilon$ and $p$  should be expressed as  functions of
$\varepsilon^\prime$ and $p^\prime$, according to the Lorentz transformation
[9]. Because the time increments (counted from the initial
preparation of the system) are given by $t^\prime$ and $t =t^\prime /\sqrt{1-V^2}$,
respectively, the relation
\begin{equation}
P^\prime (p^\prime, t^\prime) = {1+ Vv^\prime \over \sqrt{1-V^2} } P \Bigl({p^\prime+
V\varepsilon^\prime \over \sqrt{1-V^2} }, \,
{t^\prime \over \sqrt{1-V^2}} \Bigr)
\end{equation}
is required. One has $v=(v^\prime+V)/(1+ Vv^\prime)$, $\xi_t = (1-V^2)^{1/4} \xi_{t^\prime}$,
and, using the Stratonovich calculus, $dp/dt = (1+ Vv^\prime) dp^\prime/dt^\prime$,
thus the transformed stochastic kinetic (1) takes the form
\begin{equation}
{dp^\prime \over dt^\prime}=-\gamma {v^\prime + V \over (1+V v^\prime)^2}
+ {(1-V^2)^{1/4} \over
1+Vv^\prime} {\xi_{t^\prime}},
\end{equation}
where $v^\prime = p^\prime/\sqrt{{p^\prime}^2 +m^2}$.
It is easy to verify that the r.h.s.\ of (4)
solves the Fokker--Planck equation corresponding to the Eq.~(5)
if (and only if) $P(p,t)$ satisfies the proper Eq.~(2) related to the kinetic (1).
The same conclusion applies also to the general case (3).

The Eq.~(1), $\dot{p}_t= -\gamma {p/\sqrt{p^2 +m^2}} +\xi_t$, for the
momentum
or, equivalently, the (Stratonovich) equation for the velocity,
\begin{equation} \dot{v}_t= -\gamma v(1-v^2)^{3/2}/m  +\xi_t (1-v^2)^{3/2}/m,
\end{equation}
thus provides the relativistic
generalization of the diffusion Markovian description of equilibration for
noninteracting free particles system [10].
The  normalized stationary distributions are
\begin{equation}
W(p) = {1 \over 2m K_1(m\gamma/D)} e^{-(\gamma/D) \sqrt{p^2 +m^2}}
\end{equation}
\begin{equation}
{\rm and} \qquad W(v) ={(1-v^2)^{-3/2}\over 2 K_1(m\gamma/D)} e^{-(\gamma/D)m/ \sqrt{1-v^2}},
\end{equation}
where $\gamma= D\beta$. $K_n $ is a modified
Bessel function. The second moments read
\begin{equation}
\langle p^2 \rangle = {m \over \beta} {K_2(m\beta)\over K_1(m\beta)}, \quad
\langle v^2 \rangle = 1- \int_{m\beta}^\infty dz {K_0(z)\over K_1(m\beta)},
\end{equation}
so the (even) moments of the velocity, as incomplete integrals of Bessel functions,
are not given in closed  analytical form.

The nonlinear kinetic
(1) or (6) does not belong to the class of the known solvable models and
the time--dependent solution is not known. Nevertheless, there are
still the quantities, useful to characterize the relaxation
of the  nonlinear systems, which can be
expressed by  {\it quadratures} of the ({\it necessarily autonomous})
drift and
diffusion coefficients,
without solving the kinetic equation. These are the complete (over time) integrals
of (various) transient (single-time) moments [11] and stationary (two-times)
correlation functions [12]. In fact, making use of
$P(x,t|x_0)= e^{t \hat{L}} \delta(x-x_0)$, where $\hat{L}(x_0)$ is a {\it backward}
Fokker--Planck operator [8],
and computing
$X(x_0) = \int_0^\infty dt \langle x_t \rangle_{x_0} $, where
the integrand is a conditional
moment equal  $x_0$ at the initial time $t=0$,
by a formal integration $\int_0^\infty dt e^{t \hat{L}} = -\hat{L}^{-1}$ one obtains
\begin{equation}
 \hat{L} X = - x_0,
\end{equation}
which is
a nonhomogeneous linear equation of the first order for the
function $dX/dx_0$. The $\langle x \rangle_{\rm st} =0$
is required for the convergence of $X$ and the condition $X(0)=0$
determines a first constant of integration for Eq.~(10). The second one should
be chosen in such way that the correct (deterministic) result is recovered
when the noise strength $D \to 0$. The Eq.~(10) differs from
the Pontryagin equation for the mean first passage time [13]
in nonhomogeneous term
($-x_0$ instead of $-1$).
The integrals of
stationary autocorrelations are then given by
$S_x= \int_0^\infty dt \langle x_t x_0 \rangle_{\rm st} =
\int dx  W(x) X(x) x$.
The last quantity (i.e., the power spectrum
function taken at zero frequency $S_x=S_x(0)$),
is called, after some normalization (usually divided by the stationary variance)
resulting in  the proper (time) dimension,
the {\it relaxation
time} $T$ [12].  The analytical expression for this quantity has been
first obtained by Jung and Risken [14]. Then, by a similar consideration,
the formulae for the quantities of the former type, called generally
nonlinear relaxation times, have been found [15, 11].

Identifying the appropriate coefficients from Eqs.~(1) and (6),
and carrying
out the integrals of the Jung and Risken formula [14]
we obtain the exact results
\begin{equation} S_{v}=D/\gamma^2,
\end{equation}
\begin{equation}
S_{p}= Dm^2/\gamma^2 + {3mD^2 \over \gamma^3} {K_2(m\gamma/D) \over
K_1(m\gamma/D)} -D^3/\gamma^4.
\end{equation}
The same Eq.~(11) is obtained for the ordinary Brownian motion,
$S_p^o = m^2 S_v^o = D m^2/\gamma^2$.
Introducing $T_x= S_x/
\langle x^2 \rangle$ and {\it using $m$ and $m/\gamma$ as units of momentum and time}
[which is equivalent with
considering the generic forms of Eqs.~(1), (6) with
a single ({\it implicit})  parameter $D=\tau $ and $m=\gamma=1$]
one gets for {\it dimensionless} quantities
\begin{equation} T_{v}= \Biggl\{
\int_0^1 dv {\exp[(-\tau \sqrt{1-v^2})^{-1}] \over K_1(\tau^{-1})}  \Biggr\}^{-1},
\end{equation}
\begin{equation} T_{p}= {1-\tau^2+3\tau K_2(\tau^{-1})/K_1(\tau^{-1})
\over K_2(\tau^{-1})/K_1(\tau^{-1})},
\end{equation}
where $\tau$ is {\it a dimensionless temperature}, $\tau= D/m\gamma
= k_B T / mc^2$. Note that (13) is the alternative form of
$T_v = \tau/ \langle v^2 \rangle$,
where $\langle v^2 \rangle$ is given by Eq.~(9),
which is more
suitable for numerical computations.
The respective quantities of the ordinary Brownian motion
are simply equal to the inverse of the relaxation rate
of a linear model (1), $T_v^o= T_p^o = 1$, and do not depend
{\it explicitly} on temperature [16]. In contrast
both $T_v$ or $T_p$
increases with temperature,
see Fig.~1,
approaching the asymptotic
$T_v(\tau)   \approx \tau +\pi/2 $ or
$T_p (\tau) \approx (5/2)\tau$, respectively. The former asymptotic law
follows from representation (9) [the complete integral of $K_0$ is equal to $\pi/2$],
the latter is obtained by  the use of
the asymptotic formula $K_n(1/\tau) \approx (n-1)! (2\tau)^n/2$ (for $\tau
\to \infty$) [17]  in Eq.~(14). For small $\tau \to 0$
$T_v(\tau) \approx 1+ (3/2)\tau$ and
$T_p (\tau) \approx 1+3\tau$.
So, the decay of both ({\it stationary}) autocorrelations in a
relativistic case proceed slower in higher temperatures,
and slower than for the ordinary Brownian motion.
We want to stress however that, in the case of nonlinear system, the
relaxation time $T_x$ should {\it not} be interpreted
as a characteristic of
the equilibration process from  a certain {\it nonstationary} state.
One has in particular $T_x (\tau=0) =1$, which would be
misunderstood that the (deterministic) evolution of  the relativistic
and classical system
proceeds with the same (or at least similar) speed. Meanwhile it may be
generally proved that  at the limit
$\tau \to 0$ only the lowest order terms of the
drift and diffusion coefficients
contribute to $T_x$. Thus, such a conclusion applies at most to the
relaxation from the close to the equilibrium states.

In order to compare the nonstationary behavior
let us define the {\it transient relaxation time}
\begin{equation}
T_{\tau}(x_0) = x_0^{-1} \int_0^\infty dt \langle x_t \rangle_{x_0},
\end{equation}
which is related to the quantity $X$, Eq.~(10), via
$x_0 T_\tau(x_0) = X_\tau (x_0)$. Then $P_\tau(p_0)$ satisfies
$$-p_0 P_\tau^\prime /\sqrt{1+p_0^2} + \tau P_\tau^{\prime\prime} = -p_0.$$
The particular solution of the nonhomogeneous equation is
$P_\tau^\prime =\sqrt{1+p_0^2} +\tau$, whereas the (general) solution of the
homogeneous equation is singular at $\tau =0$. Thus,
\begin{equation}
T_{\tau} (p_0) = [p_0 \sqrt{1+p_0^2} +\log(p_0+ \sqrt{1+p_0^2})]/2p_0 +\tau.
\end{equation}
Similarly,
$$[-v_0(1-v_0^2)^{3/2} -3\tau v_0 (1-v_0^2)^2 ] V_\tau^\prime
 + \tau (1-v_0^2)^3 V_\tau^{\prime\prime} = -v_0$$
has a (regular at $\tau=0$) particular solution $V_\tau^\prime =(1-v_0^2)^{-3/2}$, so
\begin{equation}
T_{\tau} (v_0) =  1/\sqrt{1-v_0^2} \equiv \varepsilon_0.
\end{equation}
The deterministic (generic) Stokes equations,
$\dot{p}= -{p/\sqrt{p^2 +1}}$ and $\dot{v}= - v(1-v^2)^{3/2}$, are solvable by
quadratures. The solution, with the initial condition $x_0$, is
\begin{equation}
t(x)= t_x (x) - t_x(x_0),
\end{equation}
{\rm where} $\quad t_p(p)=-\sqrt{1+p^2}+\log|(1+\sqrt{1+p^2})/p|$,
and $t_v (v)=-(1-v^2)^{-1/2}+\log|(1+\sqrt{1-v^2})/v|$,
respectively.
One can verify that the direct calculation of
$\int_0^\infty dt x(t)/x_0 = \int_{x_0}^0 dx t(x)/x_0$  yields $T_0 (x_0)$
as given by Eq.~(16) or (17). The previous results (11) and (12) are  also
recovered when $T_\tau(x_0)$ is integrated with $x_0^2 W(x_0)$, where $W$
is a stationary probability density distribution (8) or (7), respectively.
Both (dimensionless) transient relaxation times for nonrelativistic
Brownian motion are equal to unity. The relativistic relaxation of momentum depends both on
the temperature and the initial state. $T_\tau (p_0)$, Eq.~(16), is an increasing
function of both arguments and has the following asymptotic
$T_\tau (p_0) \approx 1+ \tau +p_0^2/6$, for $p_0 \to 0$, and
$T_\tau (p_0) \approx  |p_0|/2+ \tau$, for $p_0 \to \infty$.
The result for  the velocity is somehow unexpected. The
Eq.~(17) shows that indeed the relativistic relaxation proceeds slower,
however does not depend on temperature. It turns out that with the increase of
temperature the early stage of evolution becomes faster in such a way
that compensates the subsequent slower long-time behavior, see Fig.~3.
The transient relaxation time $T_\tau(v_0)$ is equal to the (dimensionless)
initial energy of  the relativistic Brownian particle $\varepsilon_0$.
The plots of $T_0 (x_0)$ are shown in Fig.~2.

The equation (1) (in a generic form) has been solved
numerically by the Runge--Kutta method for $10^4$
sample realizations of the white Gaussian noise, for different
initial conditions and temperatures. $\langle p_t \rangle$ and
$\langle v_t \rangle$ have then been computed as the arithmetical
average of appropriate values
obtained for different trajectories of the noise.
The typical curves are plotted in Figs.~3 and 4,
together with the corresponding deterministic results (18).
The solution of the relativistic Stokes equation for velocity
changes qualitatively,
when $v_0$ exceeds $1/2$. In fact below this value of the initial state
the function $v(t)$ is convex down for all $t$   [as well as (always) $p(t)$].
However for $v_0>1/2$ at the early stage of evolution the $v(t)$ appears
convex up [in order to achieve the required
asymptote
$v(t)=1=c$ for $v_0 \to 1$]. The former {\it subrelativistic}
case, with particular $v_0 = 2/5$ ($p_0 \approx 0.436$), and the
latter, with $p_0=4$ ($v_0 \approx  0.970$), are presented.

The results of the paper may be summarized as follows. We have shown that
using the relations of  the relativistic dynamics we can generalize the
Brownian motion theory in a way consistent with thermodynamical requirements.
The general solvable  equation (10) for the quantities simply related
to the transient
relaxation times (15)
has been found. These, as well the
stationary relaxation times $T$, have been found analytically for the
relativistic  momentum and velocity (stochastic) kinetics, Eqs.~(13, 14,
16, 17).
All these quantities (considered as  dimensionless) are equal unity for the
ordinary  Brownian motion. The relaxation times $T_x (\tau)$, by the
normalization equal unity in the deterministic limit $\tau \to 0$,
exhibit positive first order
corrections for small temperatures, becoming just proportional to $\tau$ in a
high temperature asymptotic. Thus the decay of correlations in a stationary state
proceed slower than for the nonrelativistic Brownian motion.
The transient relaxation time  [describing the relaxation from a given initial
state $x_0$,
and by the normalization equal unity for the
deterministic system  set initially at the equilibrium, $T_0 (0)=1$]
of momentum $T_\tau(p_0)$ is an affinic function of temperature.
The asymptotic behavior for the large argument(s) reads
$T_\tau(p_0) \approx |p_0|/2+ \tau$. The transient relaxation time for the
velocity (17) does not depend on temperature, and it is equal to the dimensionless
initial energy of the relativistic Brownian particle.
The {\it thermal corrections}
to the considered global characteristics of the relaxation
are linear with the slope of the unity order and
appear (even) for the
noninteracting free relativistic Brownian particles system.
The
corrections can be observed for sufficiently high {\it physical}
temperatures.
For, e.g., hydrogen particle (proton)
the dimensionless quantity, which characterize the deviation from
the ordinary nonrelativistic behavior,
$\tau = k_B T / mc^2 \approx 10^{-13} T$[K].
So for the temperature $T \sim 1 {\rm MeV} \sim 10^{10}$K the {\it relative}
correction
is of the order $0.1$\%. The correction to transient relaxation time of the
velocity has the  kinematic origin only and for $T_\tau (p_0)$ both kinds
of corrections are  decoupled. Note that the general dependence on temperature
in physical units, introduced (multiplicatively) by the coefficient $\gamma(T)$,
should be counted separately [16].

The three-dimensional generalization of the proposed diffusion description
is obvious, however in contrast to the ordinary Brownian motion theory
it leads to the set of the three  coupled
($\vec{v} = \vec{p}/\varepsilon = \vec{p}/\sqrt{m^2+\vec{p}^2}$)
Langevin equations
(with statistically independent noises in each direction) or equivalently
to the (forward) Fokker--Planck description [8]
[$\nabla = \partial/ \partial \vec{p}$]
\begin{equation}
\partial_t P(\vec{p},t|\vec{p_0}) =
\gamma \nabla \circ (\vec{p} P/\varepsilon) +
D \nabla^2 P.
\end{equation}
The required form of  the (Gibbs-Boltzmann) stationary distribution is still
recovered, however the calculations of relaxation times cannot
be analytically completed in this case.

\begin{figure}
\caption{
Plots of  $T_p$ [Eq.~(14), upper curve] and
$T_v$ [Eq.~(13)] vs $\tau$ (dimensionless temperature).
The latter has been computed numerically by the Romberg method.}
\end{figure}

\begin{figure}
\caption{
Plots of $T_0(p_0)$ [Eq.~(16), $\tau=0$,  upper curve]  vs $p_0$
and $T_\tau (v_0)$  [Eq.~(17)] vs ${\bf 10 \times} v_0$. }
\end{figure}

\begin{figure}
\caption{
Plots of  $\langle v_t \rangle_{v_0}/v_0$ vs $t$. The upper and lower curves
correspond to $v_0 \approx 0.970$ and $v_0=2/5$; respectively. The deterministic
solutions, Eq.~(18), are marked by $0$ ($\tau=0$). The dotted curve, $e^{-t}$,
corresponds to the nonrelativistic universal result. The remaining curves have
been
obtained numerically for $\tau = 0.2$ and $0.6$, respectively.}
\end{figure}

\begin{figure}
\caption{
The same for $\langle p_t \rangle_{p_0}/p_0$ vs $t$.}
\end{figure}


\begin{thebibliography}{}
\bibitem{16}
All the remaining consideration concern on
dimemsionless quantities when particularly the time unit
is fixed by $m/\gamma$. Thus the changes of values of
quantities in physical units introduced by
$\gamma= \gamma(T)$  are not included in
our formulations claiming the temperature dependence.
The temperature dependence of
$\gamma=\gamma(T)$
is determined by the properties of the interaction with the thermal bath.
For the hard sphere model one has in particular $\gamma(T) \propto \sqrt{T}$
(which corresponds to the ordinary nonrelativistic result), for
$\tau \equiv k_B T / mc^2 \ll 1$ and
$\gamma(T) \propto T$, for high temperatures
$\tau \gg 1$, see Ref.~[10].
\end{thebibliography}
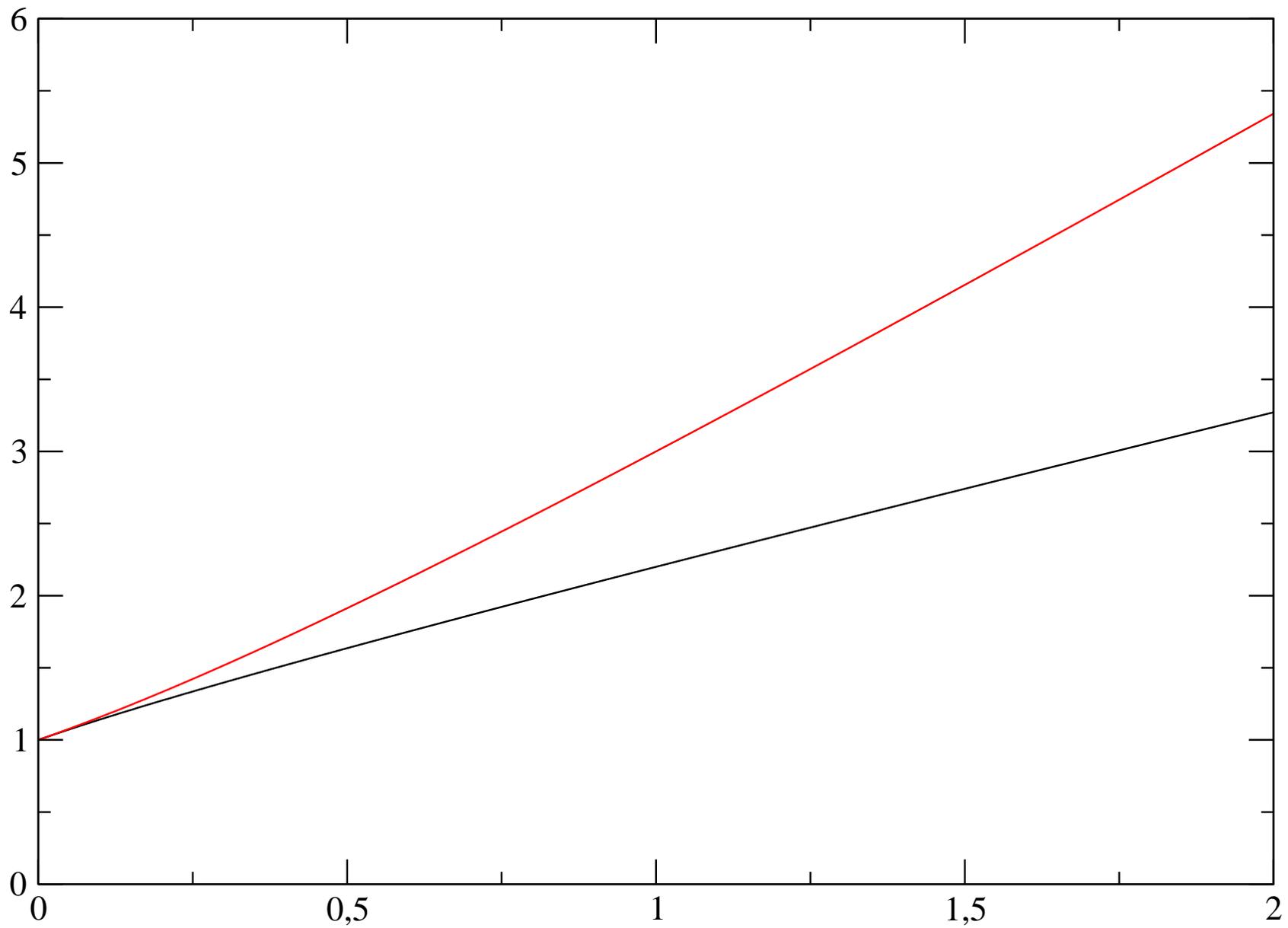
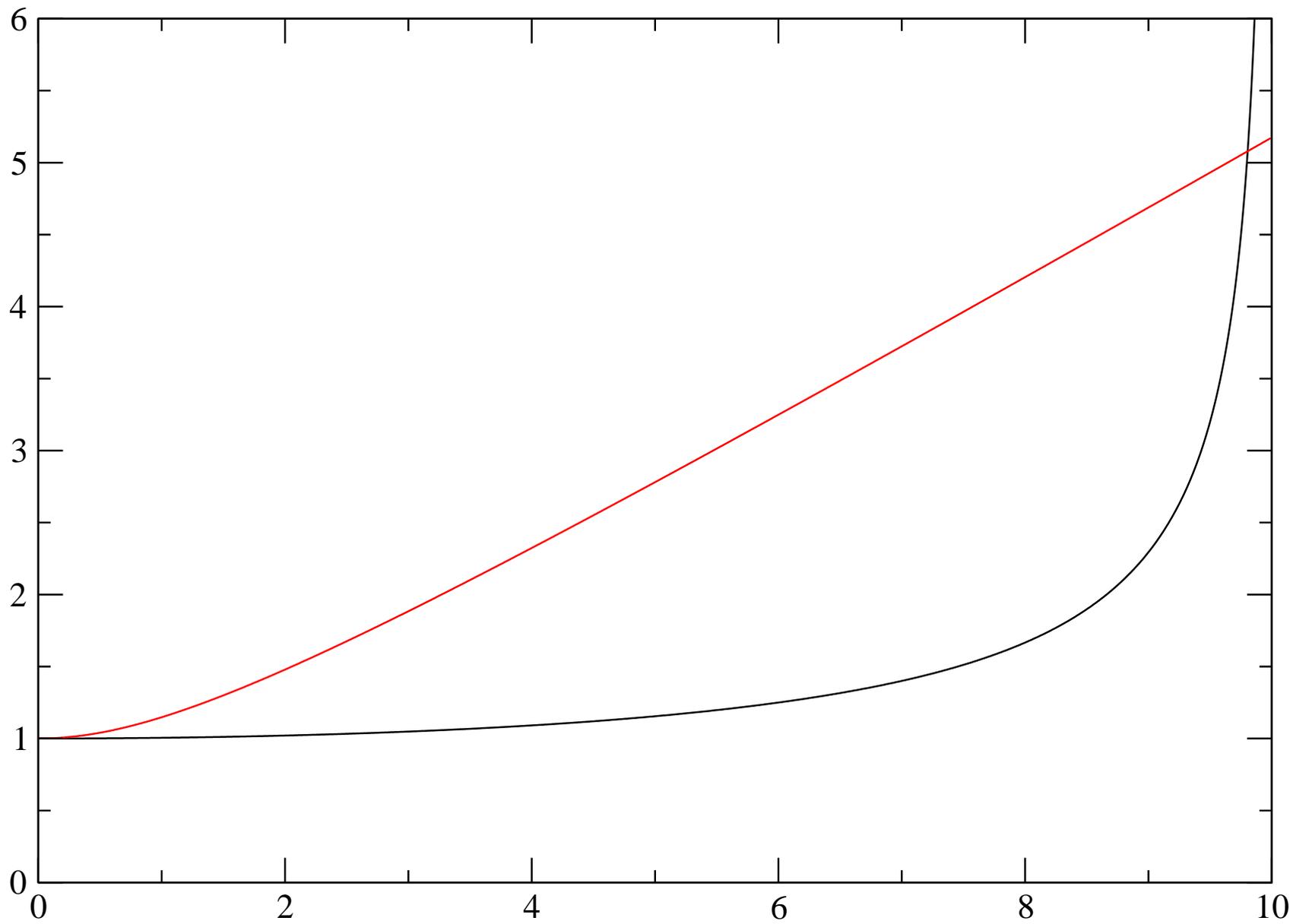
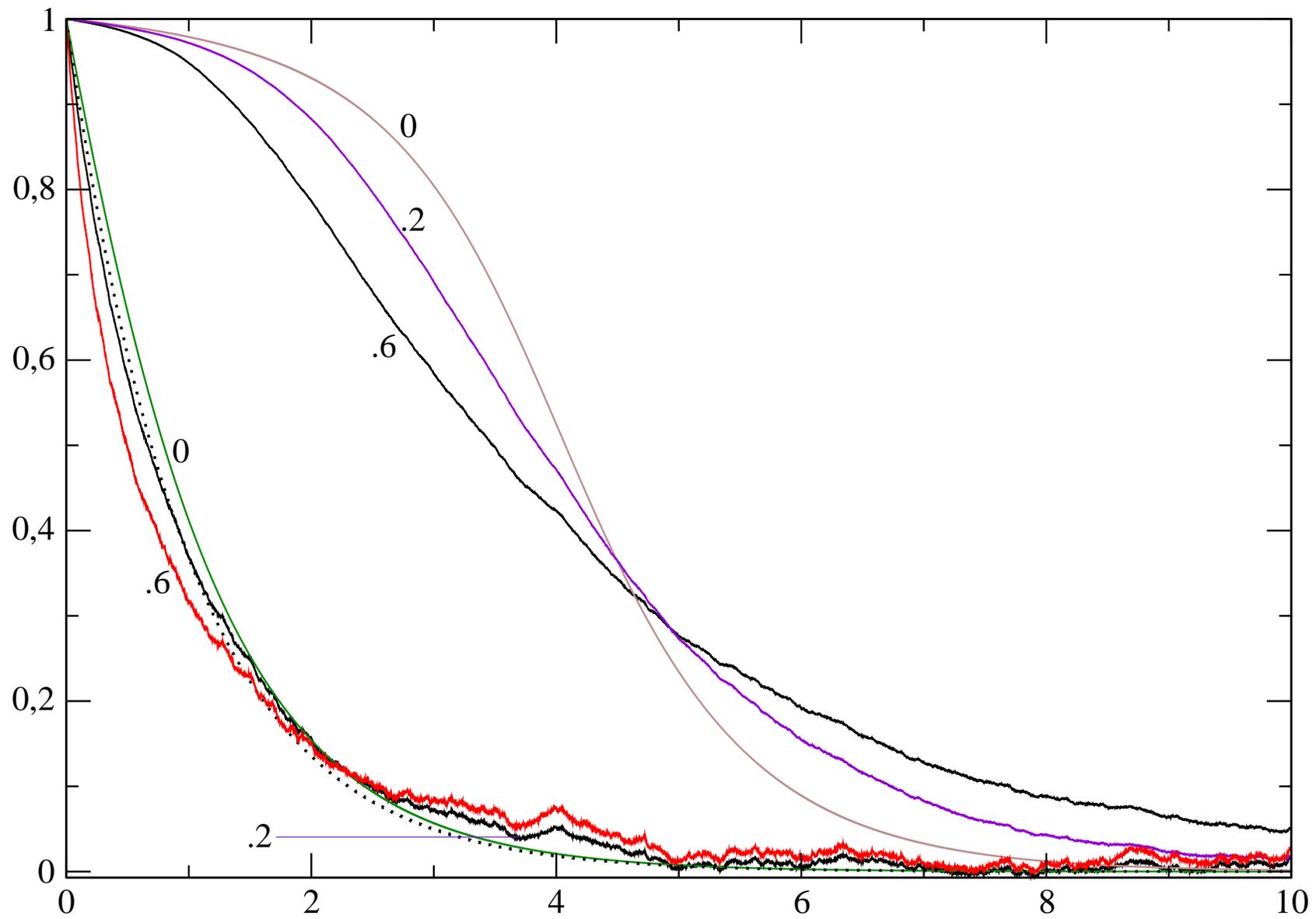
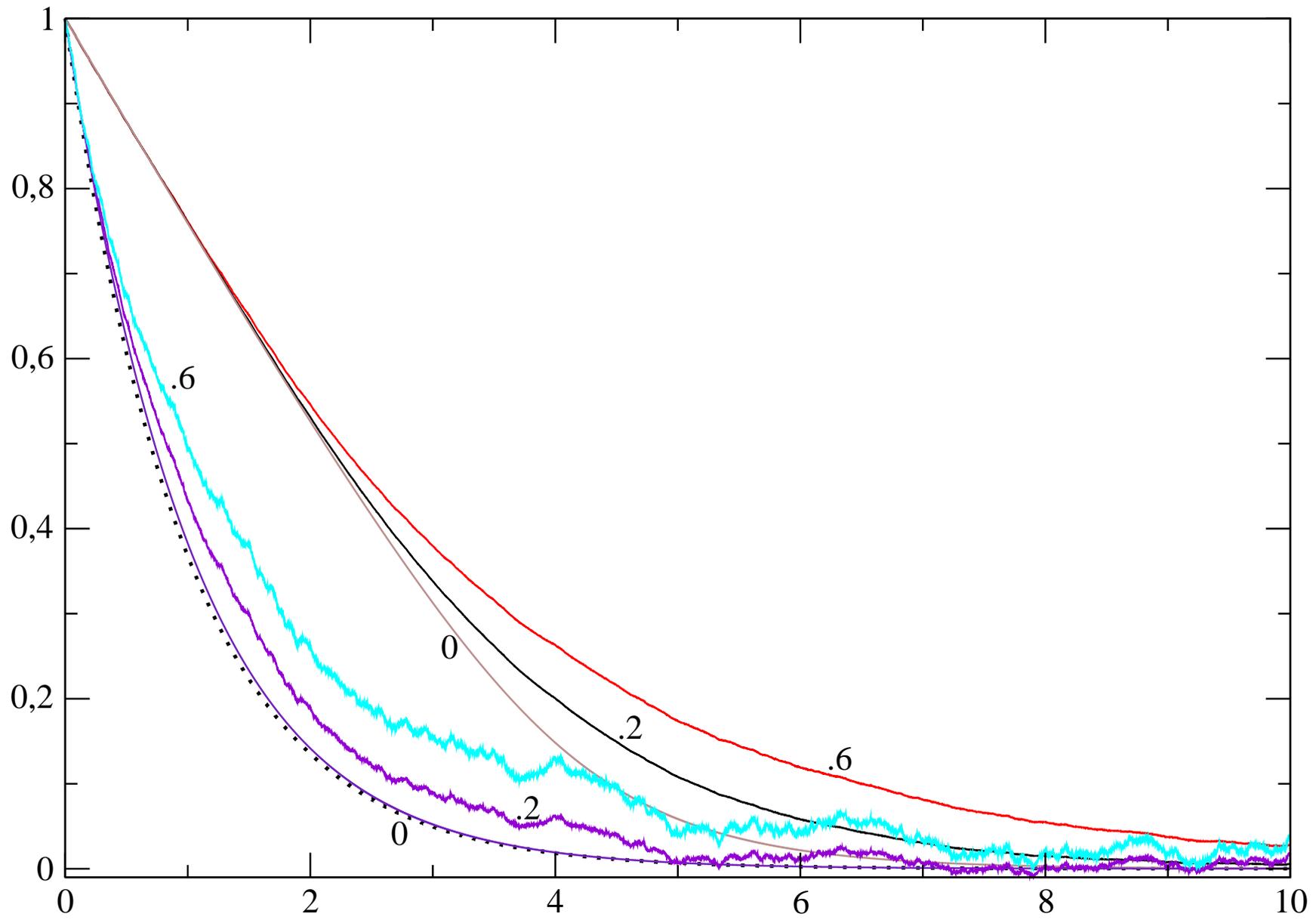
\end{document}